\documentclass[aps,prd,amssymb,eqsecnum,showpacs] {revtex4}

\begin{document}

\title{The affine-null metric formulation of Einstein's equations}

\author{ J. Winicour${}^{1,2}$
       }
\affiliation{
${}^{1}$ Department of Physics and Astronomy \\
        University of Pittsburgh, Pittsburgh, PA 15260, USA\\
${}^{2}$ Max-Planck-Institut f\" ur
         Gravitationsphysik, Albert-Einstein-Institut, \\
	 14476 Golm, Germany \\
	 }

\begin{abstract}

The details are presented of a new evolution algorithm for the characteristic 
initial-boundary value problem based upon an affine parameter rather than the
areal radial coordinate used in the Bondi-Sachs formulation.  The advantages over
the Bondi-Sachs version are discussed, with particular emphasis on the application
to the characteristic extraction of the gravitational waveform from Cauchy 
simulations of  general relativistic astrophysical systems.

\end{abstract}


\maketitle

\section{Introduction}

There has been important progress in computing accurate  gravitational waveforms
by means of Cauchy-characteristic extraction (CCE)~\cite{cce}, whereby data from a
Cauchy simulation provides the inner boundary data for a characteristic evolution
extending to future null infinity $\mathcal{I}^+$, where the waveform is defined
unambiguously. CCE has become an important tool for gravitational wave data
analysis~\cite{ninja}. It has been applied to compute waveforms from simulations
of binary black hole inspiral and mergers~\cite{reis1,reis2, mariext}, from
rotating stellar core collapse~\cite{reis3,collapsar}, to explore the memory
effect~\cite{reis4} and to study the effect of spin on gravitational waves from
precessing binary black holes~\cite{ccespin}.

A CCE module~\cite{extractool} has been prepared for public use as part of the
Einstein toolkit~\cite{einstk}. The module is based upon the PITT null
code~\cite{isaac,highp}, which implements the worldtube-nullcone
version~\cite{tam} of the Bondi-Sachs~\cite{bondi,sachs} characteristic 
initial-boundary value problem. There are technical complications in applying the
Bondi-Sachs formulation to CCE arising from the use of an areal radial coordinate
to parametrize the outgoing null geodesics.  This paper considers an alternative
approach to the worldtube-nullcone problem which replaces the areal coordinate by
an affine parameter. The details of an evolution algorithm for the affine system
of Einstein equations are presented. The comparative advantages with the
Bondi-Sachs version for application to CCE are discussed.

Recent success in simulating general relativistic astrophysical systems has been
achieved by Cauchy codes, which evolve the spacetime metric inside an artificially
constructed outer boundary. In doing so, it is common practice to compute the
gravitational waveform from data on an extraction worldtube inside the outer
boundary, using perturbative methods based upon introducing a Schwarzschild
background in the exterior region. This has been carried out using the 
Regge-Wheeler-Zerilli ~\cite{regwh,zeril} treatment of the perturbed metric, as
reviewed in~\cite{nagrez},  and also by calculating the Newman-Penrose~\cite{np}
Weyl curvature component $\Psi_4$, as first done for the binary black hole problem
in~\cite{bakcamluo,pretlet,camluomarzlo,bakcenchokopvme}. In this approach, errors
arise from the finite size of the extraction worldtube, from nonlinearities and
from gauge ambiguities involved in the arbitrary introduction of a background
metric. The gauge ambiguities might seem less severe in the case of $\Psi_4$ (vs
metric) extraction, but there are still delicate problems associated with the
choices of a preferred null tetrad and  preferred worldlines along which to
measure the waveform (see ~\cite{lehnmor} for an analysis).

In order to properly approximate the waveform at $\mathcal{I}^+$ the extraction
worldtube must be sufficiently large but at the same time causally and numerically
isolated from errors propagating in from the outer boundary. Considerable
improvement in the perturbative approach has resulted from techniques for dealing
with large outer boundaries and extrapolating the extracted waveform to infinity.
However, this is not an ideally efficient approach. It is especially impractical
in simulations of stellar collapse, where it is most strategic to restrict the
computational domain to just outside the stellar surface. CCE is a different
approach which is specifically tailored to study radiation at $\mathcal{I}^+$. 

In problems with isolated sources, the radiation zone can be compactified inside a
finite grid boundary with the metric rescaled by $1/r^2$ as an implementation of
Penrose's~\cite{penrose} conformal boundary at $\mathcal{I}^+$. Because
$\mathcal{I}^+$ is a null hypersurface, no extraneous outgoing radiation condition
or other artificial boundary condition is required.  In CCE, Cauchy data on the
extraction worldtube provides the inner boundary data for a characteristic
evolution extending to a compactified $\mathcal{I}^+$, where the waveform is
defined unambiguously by geometric methods. This eliminates waveform error due to
asymptotic approximations and gauge ambiguities  introduced by the choice of
extraction worldtube. In addition, the extraction worldtube  can be placed in the
near zone surrounding the sources in order to enhance computational efficiency.
See~\cite{winrev} for a review.

High accuracy waveforms from a binary inspiral and merger are important for the
design of the detection templates that are critical for the success of
gravitational wave astronomy. This has stimulated efforts to increase the accuracy
of characteristic evolution for use in CCE.  Another global approach applicable to
isolated systems is to base the Cauchy problem itself on the analogue of the
hyperboloidal Cauchy hypersurfaces  in Minkowski space, which asymptote to
$\mathcal{I}^+$. This approach, first extensively developed by
Friedrich~\cite{helconf}, is potentially the basis for a very attractive numerical
approach to simulate gravitational wave production. For reviews of progress on the
numerical implementation see~\cite{fraurev,saschrev1,saschrev2}. In spite of the
attractiveness of the hyperboloidal approach and its recent success with model
problems~\cite{zen,moncrinn,rinn,bardsarbuch}, considerable work remains to make
it applicable to systems of astrophysical interest. 

The Cauchy evolution codes have incorporated increasingly sophisticated numerical
techniques, such as mesh refinement, multi-domain decomposition, pseudo-spectral
collocation and high order (in some cases eighth order) finite difference
approximations. Work has begun to incorporate such techniques in characteristic
codes~\cite{hordnull}. However, such high accuracy methods cannot by themselves
cure some of the major complications and sources of error arising in CCE. The
timelike extraction worldtube ${\cal T}$ at the inner boundary of CCE is
constructed from a coordinate sphere $x^2+y^2+z^2 ={\cal R}^2$, ${\cal R}=const$, 
cut out from the Cartesian Cauchy grid. However the radial grid points of the
Bondi-Sachs system are based upon an areal coordinate $r$, with the angular grid
lying on the spheres $r=R=const$. As a result, the extraction worldtube  ${\cal
T}$ does not lie on the gridpoints of the Bondi-Sachs system (except for special
cases such as spherical symmetry). This necessitates the introduction of an
auxiliary characteristic coordinate system in the neighborhood of  ${\cal T}$ in
which the radial coordinate is replaced by an affine parameter $\lambda$ along the
outgoing null rays. By taking advantage of the affine freedom,  ${\cal T}$ can
then be parametrized by $\lambda=0$. The Cauchy data is first transformed into the
affine characteristic system and expanded about  $\lambda=0$ to a sufficient power
of $\lambda$ to determine data for the inner $r$-grid points of the Bondi-Sachs
system in the neighborhood of  ${\cal T}$. This is a complicated procedure which
introduces interpolation error and has even led to inconsistent inner boundary
conditions in the initial implementation of CCE (see~\cite{extractool} for a
discussion).

In view of this, the question naturally arises: Why not use the affine-null system
in the first place for the characteristic evolution algorithm and grid? The
history behind this choice goes far back. It has to do with the simple
hierarchical structure that the Einstein equations take in the Bondi--Sachs
system, but  which is seemingly broken in the affine system. We explain this in
Sec.~\ref{sec:nullsph}.

The difference in behavior between an areal coordinate $r$ and an affine parameter
arises from focusing effects on the null rays.  The affine coordinate $\lambda$
only becomes singular at caustics whereas the areal coordinate $r$ also becomes
singular at points where the expansion of the null rays vanish.  We deal here with
the vacuum Einstein equations, where such focusing effects do not arise in the
spherically symmetric case and the areal coordinate is also an affine parameter
along the radial null geodesics. However, there is another important application
of characteristic coordinates to cosmology where, due to the lensing effect of
matter, even in spherical symmetry the areal coordinate is not affine. The natural
role of the past null cone in astronomical observations has been incorporated into
to a new approach  to cosmology~\cite{ksachs,ellis}. Bishop and his
collaborators~\cite{bishc1,bishc2,bishc3} have initiated a program to implement
this null cone version of observational cosmology by means of a characteristic
evolution code based upon the Bondi-Sachs formalism. In principle, data obtained
from observations on the past null cone could be evolved backward in time to
obtain the earlier history of the universe.  At present, their simulations have
been confined to the spherically symmetric case but they recognized that an areal
coordinate would limit the approach to the region of the universe prior to
refocusing. They developed a stable, convergent characteristic evolution code in
which the areal coordinate was replaced by an affine parameter. The techniques
developed in the present paper could be easily generalized to include matter and
applied to extend their treatment to anisotropic cosmologies. 

In Sec~\ref{sec:nulaff},  we discuss the details of the Einstein equations in the
affine-null system and show how the hierarchical structure of the evolution system
can be restored. This provides the basis for a new worldtube-nullcone evolution
algorithm.

\section{Null spherical coordinate systems}
\label{sec:nullsph}

The coordinates of both the Bondi-Sachs system and null-affine system are based
upon a family of outgoing null hypersurfaces emanating from the spherical
cross-sections of a timelike worldtube ${\cal T}$, where the null coordinate $u$
labels these hypersurfaces and the angular coordinates $x^A$ ($A=2,3)$ label the
spherical set of null geodesic rays. In the Bondi-Sachs system, the surface area
coordinate $r$ labels the points along the outgoing null rays. In the resulting
$x^\alpha=(u,r,x^A)$ coordinates, the metric takes the Bondi-Sachs
form~\cite{bondi,sachs}
\begin{equation}
   ds^2=-\left(e^{2\beta}{V \over r} -r^2h_{AB}U^AU^B\right)du^2
        -2e^{2\beta}dudr -2r^2 h_{AB}U^Bdudx^A +r^2h_{AB}dx^Adx^B,
   \label{eq:bmet}
\end{equation}
where  $\det(h_{AB})=\det(q_{AB})=q(x^C)$, with $q_{AB}(x^C)$ some standard choice
of unit round-sphere metric. The fields $\beta$,  $U^A$, $V$ and $h_{AB}$ are
functions of $(u,r,x^A)$. Here $h_{AB}$ is the metric of the topological 2-spheres
$(u=const,r=const)$ after conformal rescaling by $1/r^2$ to surface area $4\pi$.
Its inverse is defined by $h^{AC}h_{CB}=\delta^A_B$.

The affine-null system is similarly based upon the outgoing null hypersurfaces
$u=const$ emanating from ${\cal T}$ with coordinates $x^A$ labeling the null rays
but now an affine parameter $\lambda$ is used to coordinatize points along the
rays. The affine freedom
$$\lambda\rightarrow A(u,x^A)\lambda +B(u,x^A)
$$
is used to prescribe the normalization  $(\nabla^a u)\nabla_a  \lambda =-1$ and
set $\lambda= 0$ on ${\cal T}$. In the resulting  $x^\alpha = (u, \lambda, x^A)$
coordinates, the metric takes the form 
\begin{equation}
ds^2= -({\cal V}-g_{AB} W^A W^B)du^2-2du d\lambda -2g_{AB} W^A du dx^B
       +g_{AB} dx^A dx^B
       \label{eq:naffmet}
\end{equation}
In addition, we again set $g_{AB}=r^2 h_{AB}$, where $\det h_{AB} =\det q_{AB}$
with $q_{AB}(x^C)$  a unit round-sphere metric. However, $r$ is now a metric
function of $(u,\lambda,x^A)$ along with $W^A$, ${\cal V}$ and $h_{AB}$.

At the metric level, such affine-null coordinates were introduced by
Sachs~\cite{sachsdn} in formulating a double-null initial value problem. They are
also the natural coordinates adopted in the Newman-Penrose~\cite{np} formulation
of the Einstein equations in terms of a null tetrad and the associated Weyl tensor
components.  The affine coordinate $\lambda$ is singular only at caustics whereas
the areal coordinate $r$ is also singular at points where the expansion of the
null rays vanish. In particular, this occurs at the points on a stationary event
horizon. As a result, codes based upon an areal coordinate have poor accuracy in
tracking the late time tail preceding black hole formation, cf.~\cite{close2} for
a discussion in the context of black hole perturbation theory. The affine null
metric and and the Bondi-Sachs metric are related by the transformation
$\lambda(u,r,x^A)$ determined by
\begin{equation}
       \partial_r \lambda(u,r,x^A) = e^{2\beta}.
\end{equation}       
However, the simplicity of this transformation is misleading because the surfaces
$r=const$ which determine the partial derivatives $\partial_A$ and $\partial_u$ in
the Bondi-Sachs system differ from the $\lambda=const$ surfaces that determine the
partial derivatives $\partial_A$ and $\partial_u$ in the affine-null system. It is
important to keep this distinction in mind in the following comparison of the
corresponding evolution systems.

The role of the different components of the Einstein equations in formulating a
characteristic initial value problem can be best described in terms of an
orthonormal  null tetrad $(L^a, N^a, M^a, \bar M^a)$, corresponding to the metric
decomposition 
\begin{equation}
       g_{ab}=-L_{(a }N_{b) }+M_{(a }\bar M_{b) }, \quad N^a L_a =-2,  
             \quad M^a \bar M_a =2.
       \label{eq:ntetrad}
\end{equation}
We choose $L_a = -u_a$ to be the future pointing normal to the null hypersurfaces
(so that $L^a$ is tangent to the outgoing rays)  and choose $M^a$ to be a complex 
spatial vector tangent to the null hypersurfaces. This uniquely determines $N^a$.
Then the vacuum Einstein equations $G_{ab}=0$ decompose into the main equations
\begin{eqnarray}
       L^b G_{ab}&=&0 
       \label{eq:hyp}  \\
          M^a M^b  G_{ab}&=&0
          \label{eq:ev}
\end{eqnarray}
and the supplementary equations
\begin{equation}
       N^b R_{ab}=0, \quad G_{ab}= R_{ab} -\frac{1}{2} g_{ab} R.
       \label{eq:supp}
\end{equation}

It is a consequence of the Bianchi identities that if the main equations are
 satisfied then $N^b R_{ab}$ satisfies a first order ordinary differential
 equation  along the null rays. As a result, if the main equations are satisfied
 and the supplementary equations $N^b R_{ab}=0$ are satisfied on the worldtube
 ${\cal T}$ then they will be satisfied everywhere. This result was first
 demonstrated for the Bondi-Sachs system in~\cite{bondi,sachs} but it also holds
 for the affine-null system. See~\cite{josh} for a recent discussion of the
 supplementary equations as a system of worldtube conservation laws that impose
 symmetric hyperbolic constraints on the worldtube data. In CCE, the worldtube
 data is supplied by solutions of the Einstein equations determined by the Cauchy
 evolution and it is assumed this data is consistent with the supplementary
 equations. Thus we concentrate here on the main equations.

First consider the Bondi-Sachs equations. In that case,  following the formalism
developed in~\cite{newtgc}, the main equations (\ref{eq:hyp}) take the schematic
form of hypersurface equations
\begin{eqnarray}
   \beta_{,r} &=&{\cal N}_\beta[h_{CD}]
    \label{eq:nbeta}\\
   (r^4 e^{-2\beta} h_{AB} U^B_{,r})_{,r}  &=&{\cal N}_U[h_{CD},\beta]
   \label{eq:nU} \\
    V_{,r} &=&{\cal N}_V[h_{CD},\beta,U^C] ,
   \label{eq:nv} 
\end{eqnarray}
where a ``comma'' denotes partial derivatives, e.g. $ \beta_{,r} =\partial_r
\beta$, and the main equations (\ref{eq:ev}) take the form of evolution equations
\begin{equation}
    M^A M^B  (rh_{AB,u})_{,r} = {\cal N}_h[h_{CD},\beta,U^C,V].
    \label{eq:nhu}
\end{equation}    
Here the ${\cal N}$-terms on the right hand sides of (\ref{eq:nbeta}) --
(\ref{eq:nhu})  can be calculated from the values of their arguments on a given
$u=const$ null hypersurface. Moreover, each ${\cal N}$-term only depends upon
previous members in the sequential order $[h_{CD},\beta,U^C,V]$. Because of this
hierarchical structure of the system, given $h_{AB}$ on an initial null
hypersurface $u=0$, the main equations can be integrated radially in sequential
order to determine the initial values of $\beta$, $U^A$, $V$ and $ h_{AB,u}$ at
$u=0$ in terms of their integration constants on ${\cal T}$, i.e.
\begin{equation}
      \beta|_{\cal T}, \, U^A|_{\cal T}, \, U^A_{,r}|_{\cal T}, \, 
      V|_{\cal T}, \,  h_{AB,u}|_{\cal T}.
      \label{eq:wdata}
\end{equation}
In addition, the location of the worldtube, specified by $r|_{\cal T}=R(u,x^A)$,
is another essential part of the data. After determining $ h_{AB,u}$ at $u=0$, the
hypersurface data $h_{AB}$  can be advanced to $u=\Delta u$ by a finite difference
procedure. Given the worldtube data (\ref{eq:wdata}), this procedure can be
iterated to form a worldtube-nullcone evolution algorithm. This evolution
algorithm is extremely simple and economical compared to Cauchy evolution
algorithms. It is the algorithm underlying the PITT null code.

Now consider the affine-null system, for which the main equations take the
schematic form,
\begin{eqnarray}
 r^{-1} r_{,\lambda\lambda} &=&{\cal H}_r[h_{CD}]
       \label{eq:lr}\\
   (r^4  h_{AB} W^B_{,\lambda})_{,\lambda}  &=&{\cal H}_W[h_{CD},r] 
   \label{eq:lw} \\
  \bigg( 2(r^2)_{,u}- {\cal V} (r^2)_{,\lambda} \bigg)_{,\lambda} &=&
     {\cal H}_{\cal V}[h_{CD},r,W^C]
   \label{eq:lv}   \\
       M^A M^B  (rh_{AB})_{,u\lambda} &=& {\cal H}_h[h_{CD},r,W^C,{\cal V}],
    \label{eq:lhu}
\end{eqnarray} where the ${\cal H}$-terms on the right hand sides of (\ref{eq:lr})
-- (\ref{eq:lhu})  can again be calculated from the values of their arguments on a
given $u=const$ null hypersurface. 

As in the Bondi-Sachs case, the ${\cal H}$-terms depends upon the metric functions
in sequential order, in this case in the order $[h_{CD},r,W^C,{\cal V}]$. However,
the hierarchical structure of the radial integration scheme is broken by the
appearance of the term $(r^2)_{,u}$ term on the left hand side of (\ref{eq:lv} ).
Thus (\ref{eq:lv}) is not a pure hypersurface equation and the radial integration
scheme does not produce an evolution algorithm in the same was as for the
Bondi-Sachs system. This was this reason that the affine-null formulation was not
chosen in building the PITT null code.

However, by reformulating the hypersurface equation (\ref{eq:lv}) by the
introduction of an auxiliary variable, the pure hypersurface form of the radial
integration scheme can be restored. This new formulation is described in
Sec.~\ref{sec:nulaff} after presenting the details of the main equations of the
null-affine system.

\section{The null affine evolution system}
\label{sec:nulaff}

In presenting the details of the Einstein equations for the affine-null system, we
begin with some useful formula for describing the metric (\ref{eq:naffmet}) and
its associated connection and curvature. We then proceed to describe the
construction a numerical evolution algorithm.

\subsection{Calculation of the Einstein tensor}

The contravariant components of the metric (\ref{eq:naffmet}) are given by 
\begin{equation}
    g^{u \lambda }=-1, \quad  g^{uu}=g^{u A} =0, \quad
    g^{\lambda A} =-W^A, \quad g^{\lambda\lambda}={\cal V},
    \quad g^{AB}=r^{-2} h^{AB}.
\end{equation}
It is convenient to introduce a dyad vector $m_A$ to represent the 2-metric by
\begin{equation}
      h_{AB}=m_{(A}\bar m_{B)}\,, \quad   h^{AB}=m^{(A}\bar m^{B)}\, ,
   \quad m^A = h^{AB}m_B\,   \quad h_{AB}m^A m^B=2.
\end{equation}
For that purpose, we chose the vector $M^a$ forming the null tetrad
(\ref{eq:ntetrad}) to lie tangent to the surfaces  $(u=const,r=const)$, so that it
has components $M^a=(0,0,M^A)$.  We  then set $M^A=r^{-1}m^A$. Here we raise and
lower indices of 2-dimensional vector and tensor fields on the sphere with
$h_{AB}$ and $h^{AB}$, e.g. $W_A=h_{AB}W^B$. We recall that $\det h_{AB} =\det
q_{AB}$, where $q_{AB}(x^C)$ is some standard choice of unit round-sphere metric.
The determinant condition implies
\begin{equation}
         h^{AB}h_{AB,u}=m^A \bar m^B h_{AB,u} =0 , 
         \quad   h^{AB}h_{AB,\lambda}=m^A \bar m^B h_{AB,\lambda} =0 .
         \label{eq:detcon}
\end{equation}
We fix the spin rotation freedom in the dyad $m^A \rightarrow
e^{i\varphi(u,\lambda,x^C)} m^A$  by requiring
\begin{equation}
             \bar m^A m_{A,\lambda}=0
             \label{eq:ml}
\end{equation}
and 
\begin{equation}
             \bar m^A m_{A,u}|_{\lambda=0}=0.
\end{equation}
Note that the determinant condition (\ref{eq:detcon}) also implies
\begin{equation}
    \bar m^A m_{A,\lambda} + m^A \bar m_{A,\lambda}=0.
      \label{eq:detlam}
\end{equation}
The spin rotation freedom then reduces to a phase factor $\varphi(x^A)$, which is
determined by the choice of conventions at $(u=0,\lambda=0)$. Given these
conventions, $h_{AB}$ and $m_A$ are in one-to-one correspondence.

As an example, for stereographic coordinates $x^A=(\eta,\rho)$ on the unit
round-sphere with metric
\begin{equation}
         q_{AB} dx^A dx^B = \sqrt{q}( d\eta^2+d\rho^2),
        \quad \sqrt {q}= \frac{4}{1+\eta^2+\rho^2},
\end{equation}
the rescaled metric on the general curved topological sphere can be represented as
\begin{equation}
    h_{AB}= \sqrt{q}\left( 
\begin{array}{cc}
    e^{2\gamma} \cosh2 \alpha & \sinh 2 \alpha \\
    \sinh 2\alpha& e^{-2\gamma} \cosh 2\alpha                 
\end{array}
    \right)  ,
\label{eq:dgamalph}
\end{equation}
\begin{equation}
    h^{AB}= \frac{1}{\sqrt{q}}\left( 
\begin{array}{cc}
    e^{-2\gamma} \cosh2 \alpha & -\sinh 2 \alpha \\
    -\sinh 2\alpha& e^{2\gamma} \cosh 2\alpha                 
\end{array}
    \right)  .
\label{eq:ugamalph}
\end{equation}
where $\gamma$ and $\alpha$ represent the two degrees of freedom. A specific
choice of polarization dyad associated with this representation is
\begin{equation}
      m_A= q^{1/4}\bigg ( e^{\gamma}(\cosh\alpha+i \sinh\alpha), 
             i e^{-\gamma}(\cosh\alpha-i \sinh\alpha)\bigg ) ,
\end{equation}
\begin{equation}
      m^A= q^{-1/4}\bigg ( e^{-\gamma}(\cosh\alpha-i \sinh\alpha), 
             i e^{\gamma}(\cosh\alpha+i \sinh\alpha)\bigg ) .
\end{equation}

The components of the Einstein tensor can be calculated in terms of the metric
functions $h_{AB}, \, r, \, W^A, \, {\cal V}$ from the components of the
curvature  tensor
\begin{equation}
         {R^a}_{bcd}=\partial_c {\Gamma^a}_{bd} - \partial_d {\Gamma^a}_{bc} 
     + {\Gamma^a}_{ce}{\Gamma^e}_{bd}-{\Gamma^a}_{de}{\Gamma^e}_{bc} \, ,
          \quad   R_{ab}={R^c}_{acb},
\end{equation}
where
\begin{equation}
         \Gamma^{a}_{bc}= \frac{1}{2}g^{ad}(\partial_b g_{cd}+\partial_c g_{bd}
             -\partial_d g_{bc})
\end{equation}
are the Christoffel symbols.
The components of the Christoffel symbols in terms of the metric
functions are given in Appendix \ref{sec:app1}.
We denote 2-dimensional covariant derivatives of tensor fields
on the sphere with respect to $h_{AB}$ by
a ``colon'', e.g.
\begin{equation}
               {W^A}_{:B}=\partial_B W^A + {}^{(h)} \Gamma^{A}_{BC} W^C
\end{equation}
where
\begin{equation}
                {}^{(h)} \Gamma^{A}_{BC}
                   =\frac{1}{2} h^{AD}(\partial_B h_{CD}
                +\partial_C h_{BD}-\partial_D h_{BC})
\end{equation}
is the Christoffel symbol associated with $h_{AB}$.

In terms of these conventions and notation, the main affine-null equations
(\ref{eq:lr}) -- (\ref{eq:lhu}) have the specific form
\begin{equation}
      0= R_\lambda^b u_{,b}  = \frac{2r_{,\lambda\lambda}}{r}
              - \frac{1}{4} h^{CD}_{,\lambda} h_{CD,\lambda} 
                 \label{eq:ar}
\end{equation}
\begin{equation}
         0=  R_A^b u_{,b} =  -\frac{1}{2r^2}(r^4 h_{AB}W^B_{,\lambda})_{,\lambda} 
                 + (\frac{r_{,\lambda}}{r})_{,A}
                 - (\frac{r_{,B}}{r})h^{BC}h_{AC,\lambda}               
                 - \frac{1}{2} h^{BC} h_{AC,\lambda:B}
                    \label{eq:aA}
\end{equation}
\begin{equation}
       0=    h^{AB}R_{AB} = {}^{(2)} {\cal R}
                      +\bigg (2 (r^2)_{,u} -{\cal V}(r^2)_{,\lambda} \bigg)_{,\lambda} 
                      +\bigg( \frac{1}{r^2}(r^4W^A)_{,\lambda} \bigg)_{:A}
                      -(\ln r^2)^{:A}{}_{:A} 
                      -\frac{r^4}{2}h_{AB} W^A_{,\lambda} W^B_{,\lambda}
                         \label{eq:av}
\end{equation}
\begin{eqnarray}
         0=  m^A m^B R_{AB} &=&m^A m^B \{ r(rh_{AB})_{,u\lambda}
                      -\frac{1}{2}(r^2 {\cal V} h_{AB,\lambda})_{,\lambda}
                     +rr_{:C}W^C h_{AB,\lambda}+(r^2W_{A:B})_{,\lambda}  \nonumber \\
                     &+&\frac{r^2}{4} h_{AB,\lambda}\bar m_C m_D(W^{C:D}-W^{D:C})
                     +\frac{r^2}{2}W^C h_{AB,\lambda:C}
                     -\frac{r^4}{2}h_{AC}h_{BD}W^C_{,\lambda}W^D_{,\lambda} \}.
                        \label{eq:aev}
\end{eqnarray}

\subsection{Restoration of the hypersurface equation hierarchy}
\label{sec:restor}

The strategy now is to use the auxiliary variable 
\begin{equation}
              {\cal Y}= {\cal V}- \frac {2r_{,u}}{r_{,\lambda}}
\end{equation}
to eliminate the explicit appearance of the $r_{,u}$ derivative in (\ref{eq:aru})
and reexpress it as a hypersurface equation for ${\cal Y}$. In that process,
substitution of  ${\cal Y}$ for ${\cal V}$ in the evolution equation 
(\ref{eq:aev}) leads to the intermediate expression 
\begin{equation}
   m^A m^B \{ r(rh_{AB})_{,u\lambda}  -\frac{1}{2}(r^2 {\cal V} h_{AB,\lambda})_{,\lambda} \}
       =  r m^A m^B \{ r h_{AB,u} -\frac {r r_{,u}}{r_{,\lambda}} h_{AB,\lambda}\}_{,\lambda}
         -  m^A m^B \frac{1}{2}(r^2 {\cal Y} h_{AB,\lambda})_{,\lambda} 
                        \label{eq:aev1}
\end{equation}
or, using (\ref{eq:ml}),
\begin{eqnarray}
    m^A m^B \{ r(rh_{AB})_{,u\lambda}  -\frac{1}{2}(r^2 {\cal V} h_{AB,\lambda})_{,\lambda} \}
      &=&  r  \{m^A m^Br(  h_{AB,u} -\frac { r_{,u}}{r_{,\lambda}} h_{AB,\lambda})\}_{,\lambda}
           -  m^A m^B \frac{1}{2}(r^2 {\cal Y} h_{AB,\lambda})_{,\lambda}.§
                        \label{eq:aev2}
\end{eqnarray}

When (\ref{eq:aev2})is inserted back into (\ref{eq:aev}) it gives a
hypersurface equation for a combination of the time derivatives $m^A m^B h_{AB,u}$
and $r_{,u}$. In order to complete an evolution system, we need an additional
radial equation for $r_{,u}$. This is obtained from the $u$-derivative of the
Raychauduri equation (\ref{eq:ar}), which determines the rate of change of the
expansion of the outgoing rays. In order to formulate a hierarchical radial
integration scheme, we introduce the auxiliary variables

\begin{equation}
      \rho = r_{,u} .
      \label{eq:rho}
\end{equation}
and
\begin{equation}
  k_{AB}= h_{AB,u} .
      \label{eq:kab}
\end{equation}
Then the $u$-derivative of (\ref{eq:ar}) gives
\begin{equation}
          \rho_{,\lambda\lambda}=\frac{\rho}{8}  h^{CD}_{,\lambda} h_{CD,\lambda} 
                 +\frac{r}{4} h^{BC}_{,\lambda}k_{BC,\lambda },
                 \label{eq:aru}
\end{equation}
where the determinant condition implies that the undifferentiated $k_{AB}$ terms
vanish, i.e.
\begin{equation}
      h^{BD} h^{CE}_{,\lambda}   h_{ED,\lambda} k_{BC} =0.
 \end{equation}  
       
In assembling the foregoing results into a radial integration hierarchy,
simplifications  result from using the spin-weighted scalars
\begin{equation}
          \sigma =  \frac{1}{4}m^A m^B h_{AB,\lambda}
     =\frac{1}{2} m^A  m_{A,\lambda} \, ,
          \label{eq:sigma}
\end{equation}
\begin{equation}
         \kappa= \frac{1}{4} m^A m^B k_{AB}, 
      \quad k_{AB}=\kappa\bar m_A \bar m_B +\bar \kappa m_A  m_B
\end{equation}
and
\begin{equation}
         J =  4( r_{,\lambda} \kappa -\rho\sigma ) 
         =r_{,\lambda} m^A m^B h_{AB,u}  -  r_{,u} m^A m^B h_{AB,\lambda} .
         \label{eq:J}
\end{equation}
Then  (\ref{eq:aev2}) becomes
\begin{eqnarray}
      m^A m^B \{ r(rh_{AB})_{,u\lambda} 
      -\frac{1}{2}(r^2 {\cal V} h_{AB,\lambda})_{,\lambda} \}
            &=&  r\bigg  (\frac { rJ}{r_{,\lambda}} \bigg )_{,\lambda}
           -  m^A m^B \frac{1}{2}(r^2 {\cal Y} h_{AB,\lambda})_{,\lambda}
                        \label{eq:aev3}
\end{eqnarray}
and (\ref{eq:aru}) becomes
\begin{eqnarray}
          \rho_{,\lambda\lambda}&=&-\rho \sigma \bar\sigma  
           -r\bar  \sigma  \kappa_{,\lambda} - r \sigma \bar \kappa_{,\lambda} 
           \nonumber \\
         &=&-\frac{\rho}{r_{,\lambda}}  (r \sigma \bar\sigma)_{,\lambda} 
                -2r \sigma \bar\sigma( \frac{\rho}{r_{,\lambda}})_{,\lambda} 
                   - \frac{r \bar  \sigma}{4} (\frac { J}{r_{,\lambda}})_{,\lambda}
                      - \frac{ r \sigma}{4}  (\frac {\bar J}{r_{,\lambda}})_{,\lambda} 
                 \label{eq:aru2}
\end{eqnarray}
where, from  (\ref{eq:ar}) and(\ref{eq:sigma}), 
\begin{equation}
        r \sigma \bar\sigma = -   r_{,\lambda\lambda}.
\end{equation}

\subsection{The evolution algorithm}
\label{sec:evalg}

By assembling the results in Sec.~\ref{sec:restor}, the main equations
(\ref{eq:ar}) - (\ref{eq:aev}), along with (\ref{eq:aru2}), now take the desired
form
desired form
\begin{eqnarray}
 r^{-1} r_{,\lambda\lambda} &=&-\sigma \bar\sigma=H_r[h_{CD}]
       \label{eq:er}\\
   (r^4  h_{AB} W^B_{,\lambda})_{,\lambda}  
             &=&2r^2  (\frac{r_{,\lambda}}{r})_{,A}  -2r r_{,B}h^{BC}h_{AC,\lambda}               
                 - r^2 h^{BC} h_{AC,\lambda:B} \nonumber  \\
             &=&    H_W[h_{CD},r]
   \label{eq:ew} \\
  \bigg({\cal Y} (r^2)_{,\lambda} \bigg)_{,\lambda} &=&{}^{(2)} {\cal R}
                      +\bigg( \frac{1}{r^2}(r^4W^A)_{,\lambda} \bigg)_{:A}
                      -(\ln r^2)^{:A}{}_{:A} 
                      -\frac{r^4}{2}h_{AB} W^A_{,\lambda} W^B_{,\lambda} \nonumber \\
                      &=&H_{\cal Y}[h_{CD},r,W^C] 
   \label{eq:ey}   \\
    \bigg  (\frac{rJ}{r_{,\lambda}}\bigg )_{,\lambda} &=&m^A m^B \{
                      \frac{1}{2r}(r^2 {\cal Y} h_{AB,\lambda})_{,\lambda}
                     - r_{:C}W^C h_{AB,\lambda}- \frac{1}{r}(r^2W_{A:B})_{,\lambda}  \nonumber    \\
                    & - &   \frac{r}{4} h_{AB,\lambda}\bar m_C m_D(W^{C:D}-W^{D:C}) 
                    -\frac{r}{2}W^C h_{AB,\lambda:C}
                     +\frac{r^3}{2}h_{AC}h_{BD}W^C_{,\lambda}W^D_{,\lambda} \} \nonumber  \\
                     &=& H_J[h_{CD},r,W^C,{\cal Y}]
    \label{eq:ej} \\
    \bigg(  \frac{ \rho}{r_{,\lambda}}\bigg)_{.\lambda\lambda}&=&
         -\frac{r}{4 r_{,\lambda}} \bigg( \bar \sigma( \frac {J}{r_{,\lambda}})_{,\lambda} 
                          + \sigma( \frac {\bar J}{r_{,\lambda}})_{,\lambda} \bigg ) 
         =H_{\rho}[h_{CD},r,J] \, ,
         \label{eq:erho}
\end{eqnarray}
where the $H$-terms can be calculated from the values of their arguments on a
$u=const$ null hypersurface.  Given $h_{AB}(u,\lambda,x^C)$, the system
(\ref{eq:er}) -- (\ref{eq:erho}) forms a hierarchy of radial hypersurface
equations which can be integrated to determine the remaining variables in the
sequential order $[r, W^A, {\cal Y},J,\rho]$.

This gives rise to the following initial-boundary  value problem. Specify the
initial hypersurface data 
\begin{equation}
              [ h_{AB},r] \, ,  \quad u=0, \quad \lambda \ge 0 
\end{equation}
and the initial boundary data
\begin{equation}
   [ r_{,\lambda}, W^A, W^A_{,\lambda}, {\cal Y},J,\rho, \rho_{,\lambda} ]\, , 
     \quad u=0, \quad \lambda=0,
\end{equation}
subject to the constraint (\ref{eq:er}), which constitutes an ordinary
differential radial equation.  On the boundary, specify
\begin{equation}
             [ W^A, W^A_{,\lambda}, {\cal Y},J,\rho, \rho_{,\lambda} ]\, ,
     \quad u>0, \quad \lambda=0,
\end{equation}
subject to the conditions (\ref{eq:supp}).
Using the initial data, integrate (\ref{eq:er}) -- (\ref{eq:erho})  to determine
the initial values of  $[r,W^A, {\cal Y}, J ,\rho]$. 

Given this initialization and the boundary data, the evolution system can be
integrated by a finite difference approximation. The initial values
$\rho(0,\lambda,x^A)$ and $J(0,\lambda,x^A)$ determine the values of $r$ and
$h_{AB}$ at $u=\Delta u$ through (\ref{eq:rho}) and (\ref{eq:J}). Using the
boundary data, (\ref{eq:ew}) -- (\ref{eq:erho}) can then be integrated in
sequential order to determine   $[W^A, {\cal Y}, J, \rho]$  at $u=\Delta u$. Now
$[h_{AB},r,W^A, {\cal Y}, J, \rho]$  are known at $u=\Delta u$ and this process
can be repeated to provide a finite difference evolution algorithm. If the
algorithm converges as $\Delta u \rightarrow 0$ then it produces a solution to the
affine-null initial-boundary value problem for Einstein's equations.

\section{Discussion}

We have constructed an evolution algorithm based upon the affine-null system
which, like the Bondi-Sachs system, is based upon a hierarchy of radial equations
along the outgoing characteristics. It has the additional advantage of the
flexibility in describing an arbitrary inner worldtube boundary as a coordinate
surface. This is especially important for application to CCE, where the inner
boundary, which is constructed in terms of the Cauchy coordinates,  generically
differs from the $r=const$ Bondi-Sachs worldtubes. As a result,  the affine-null
algorithm offers the possibility of increased economy and accuracy. 

The formal solution of the null-affine problem constructed in Sec.~\ref{sec:evalg}
yields an exact solution provided the finite difference approximation converges as
$\Delta u \rightarrow 0$. A necessary condition for this is the  well-posedness of
the underlying analytic initial-boundary value problem. Well-posedness, i.e. the
existence of a unique solution which depends continuously on the data, is a
necessary condition for a successful numerical treatment. Although characteristic
evolution codes based upon the Bondi-Sachs formalism have been demonstrated to be
stable in a large number of test cases~\cite{highp,wobb}, there remains some
lingering doubt because well-posedness of the analytic problem has not yet been
established. Rendall~\cite{rend} has shown that the affine-null problem is
well-posed in the double null case where the inner boundary is also a null
hypersurface. However, Rendall's approach cannot be applied to the corresponding
problem where the inner boundary is a timelike worldtube. The well-posedness of
the worldtube-nullcone characteristic initial-boundary value problem for
Einstein's equations remains an outstanding issue.

The only source of error in CCE which does not decrease with numerical resolution
arises from the mismatch between the initial Cauchy data and initial
characteristic data. This results because the radius ${\cal R}_2$ of the outer
Cauchy boundary is larger than the radius ${\cal R}_1$ of the inner worldtube
boundary of the characteristic evolution. Whereas the Cauchy data in the region
${\cal R}_1 \le {\cal R} \le {\cal R}_2$ is chosen, say, by some constraint solver
for binary black hole initial data, the characteristic initial data for ${\cal R}
\ge {\cal R}_1$ is chosen to suppress the  initial radiation content by requiring
that the Newman-Penrose Weyl component $\Psi_0 =0$. This mismatch between the
initial Cauchy and characteristic data leads to an extraneous error in the
extracted waveform which is related to the spurious radiation content in the
Cauchy data. This error decreases as ${\cal R}_2 \rightarrow {\cal R}_1$ but
present day Cauchy codes require that the outer boundary be in the far zone of a
binary black hole to avoid incoming radiation generated by the outer boundary
condition. As a result, the full potential of CCE is not realized.

This mismatch can be eliminated by Cauchy-characteristic matching
(CCM)~\cite{ccm}. CCE is one of the pieces of CCM in which the characteristic
worldtube data is extracted from the Cauchy evolution. In CCM, data on the outer
Cauchy boundary is in turn obtained from the characteristic evolution. In doing
so, it is possible to place the radius  ${\cal R}_2$ of the Cauchy boundary just
outside the radius ${\cal R}_1$ of the characteristic extraction worldtube. In
fact, in a finite difference implementation of CCM for a model scalar wave
problem~\cite{holvorc2}, it has been possible to arrange that ${\cal R}_2
\rightarrow {\cal R}_1$ in the continuum limit. This resulted in a seamless
interface between the Cauchy and characteristic evolutions with no  mismatch in
the initial data.

The success of CCM depends upon the proper mathematical and computational
treatment of the initial-boundary value problem (IBVP) for the Cauchy evolution. 
At present, the only successful 3D application of CCM in general relativity has
been to the linearized matching problem between a characteristic code and a Cauchy
code based upon harmonic coordinates~\cite{harl}. Considerable work remains to
apply it to astrophysical systems.  The linearized harmonic code satisfied a
well-posed initial-boundary value problem, which seems to be a critical missing
ingredient in earlier attempts at CCM in general relativity. More recently, a
well-posed initial-boundary value problem has been established for fully nonlinear
harmonic evolution~\cite{wpgs,isol}, which should facilitate the extension of CCM
to the nonlinear case.

\begin{acknowledgments}

I have benefited from discussions with B. Szil\'{a}gyi on the requirements for
applying CCE to a spectral code.  The research was supported by NSF grants
PHY-0854623 and PHY-1201276 to the University of Pittsburgh.

\end{acknowledgments}

\appendix

\section{Christoffel symbols}
\label{sec:app1}

The calculation of the Ricci tensor 
\begin{equation}
   R_{\alpha\beta}=\Gamma^\nu_{\alpha\beta,\nu}-\Gamma^\nu_{\nu \alpha,\beta}
        +\Gamma^\rho_{\alpha\beta}\Gamma^\nu_{\rho\nu}
        -\Gamma^\rho_{\nu \alpha}\Gamma^\nu_{\rho \beta}
\end{equation}
which enter the main equations can be carried out explicitly in terms of the 
Christoffel symbols for the affine-null metric (\ref{eq:naffmet}). The components
are listed according to the notation $(x^0,x^1,x^A) = (u,\lambda,x^A)$.

\begin{equation}
   \Gamma^\alpha_{\alpha\nu} =\partial_\nu \ln (r^2\sqrt {q})
\end{equation}  

\begin{equation}
   \Gamma^0_{1\alpha} =0
\end{equation}  

\begin{equation}
   \Gamma^\alpha_{11} =0
\end{equation} 

\begin{equation}
   \Gamma^0_{AB} = r r_{,\lambda}h_{AB} +\frac{r^2}{2} h_{AB,\lambda}
\end{equation}  

\begin{equation}
   \Gamma^1_{AB} =\frac{r^2}{2} (h_{AC} W^C_{:B}+h_{BC} W^C_{:A}+ h_{AB,u}
              -{\cal V} h_{AB,\lambda}) +r h_{AB} (r_{,C}W^C+r_{,u} -{\cal V}  r_{,\lambda})
\end{equation}  

\begin{equation}
   \Gamma^C_{AB} =r r_{,\lambda}W^C h_{AB} +\frac{1}{2}r^2 W^C h_{AB,\lambda}
   +\frac{1}{r}( r_{,B} \delta^C_A + r_{,A} \delta^C_B -r_{,D}h^{CD}h_{AB} )
         + {}^{(h)} \Gamma^{C}_{AB}
\end{equation}  

\begin{equation}
   \Gamma^0_{0A} =-\frac{r^2}{2}( h_{AB,\lambda} W^B + h_{AB} W^B_{,\lambda})
       -r r_{,\lambda} h_{AB} W^B
\end{equation}  

\begin{equation}
   \Gamma^1_{10}  = \frac{1}{2}{\cal V} _{,\lambda} -\frac {r^2}{2} h_{AB} W^A W^B_{,\lambda}
\end{equation}  

\begin{equation}
   \Gamma^1_{1A}  =\frac {r^2}{2} h_{AB} W^B_{,\lambda}
\end{equation}  

\begin{equation}
   \Gamma^0_{00}  =- \frac{1}{2}{\cal V}_{,\lambda} +\frac {1}{2}(r^2 h_{AB} W^A W^B)_{,\lambda}
\end{equation}  

\begin{equation}
   \Gamma^C_{A1} =\frac{1}{r} r_{,\lambda} \delta^C_A + \frac{1}{2}h^{CD}h_{AD,\lambda} 
\end{equation}  

\begin{equation}
   \Gamma^1_{00}  = \frac{1}{2}W_{,0} +\frac {1}{2}(r^2 h_{AB} )_{,0} W^A W^B
      +\frac{1}{2}{\cal V} ({\cal V} -r^2  h_{AB}  W^A W^B)_{,\lambda}
      -\frac{1}{2}W^A ({\cal V} -r^2  h_{BC}  W^B W^C)_{,A}
\end{equation}  

\begin{equation}
   \Gamma^1_{0A}  = \frac{1}{2}W_{,A} +\frac {1}{2}{\cal V} (r^2 h_{AB} W^B)_{,\lambda}
     -r r_{,B}W^B h_{AC}W^C -\frac{r^2}{2}W^B ( h_{BC} W^C_{:A}+h_{AC} W^C_{:B})
       -\frac{1}{2}W^B(r^2 h_{AB})_{,0}
\end{equation}  

\begin{equation}
   \Gamma^B_{0A}  = -\frac{1}{2}W^B(r^2h_{AC}W^C)_{,\lambda} 
     +\frac{1}{2} g^{BC}(r^2h_{AC})_{,0} -\frac{r_{,A}}{r}W^B -\frac{1}{2}W^B_{:A}
       +\frac{r_{,D}}{r}h^{BD}h_{AC}W^C +\frac{1}{2}h^{BC}h_{AD}W^D_{:C} 
\end{equation} 

\begin{equation}
   \Gamma^A_{01}  = -\frac{r_{,\lambda}}{r}W^A-\frac{1}{2} h^{AB}(h_{BC}W^C)_{.\lambda}
\end{equation} 

\begin{eqnarray}
   \Gamma^A_{00}  &=& -\frac{1}{2}W^A{\cal V}_{,\lambda} 
     +\frac{1}{2} W^A(r^2 h_{BC} W^B W^C)_{,\lambda} -2W^A \frac{r_{,u}}{r}
     -h^{AB}(h_{BC}W^C)_{,u} +\frac{1}{2r^2}h^{AB}{\cal V}_{,B}  \nonumber \\
   &  - & \frac{r_{,B}}{r}h^{AB}h_{CD}W^C W^D
      -\frac{1}{2}h^{AB}(h_{CD}W^C W^D)_{,B} .
\end{eqnarray}


\begin{thebibliography}{40}

\bibitem{cce} ``Cauchy-characteristic extraction in numerical relativity'',
N.T. Bishop, R. G\'{o}mez, L. Lehner, and J. Winicour, {\em Phys. Rev. D}
{\bf 54} 6153 (1996).

\bibitem{ninja} ``The NINJA-2 catalog of hybrid post-Newtonian/numerical-relativity
waveforms for non-precessing black-hole binaries'',
P. Ajith et al: NINJA project, 
{\em Class. Quantum Grav.} {\bf 29} 124001 (2012).

\bibitem{reis1} ``Unambiguous determination of gravitational waveforms from
binary black hole mergers'',
C. Reisswig, N.~T. Bishop, D. Pollney and B. Szil\'{a}gyi, 
{\em Phys. Rev. Lett.}  {\bf  103},  221101  (2009).

\bibitem{reis2} ``Characteristic extraction in numerical relativity: binary
black hole merger waveforms at null infinity'',
C. Reisswig, N.~T. Bishop, D. Pollney and B. Szil\'{a}gyi, 
{\em Class. Quant. Grav.} {\bf 27}, 075014 (2010).

\bibitem{mariext} ``Binary black hole waveform extraction at null infinity'',
M.~C. Babiuc, J. Winicour and Y. Zlochower,
{\em Class.Quant.Grav.} {\bf 28}, 134006 (2011).

\bibitem{reis3} ``Gravitational wave extraction in simulations of rotating
stellar core collapse'',
C. Reisswig, C.~D. Ott, U. Sperhake and E. Schnetter, 
{\em Phys. Rev. D} {\bf 83}, 064008 (2011).

\bibitem{collapsar} ``Dynamics and gravitational wave signature of collapsar formation'',
 C.~D. Ott, C. Reisswig, E. Schnetter, E. O'Connor, U. Sperhake. F. L\"offler,
P. Diener, E. Abdikamalov, I. Hawke and A. Burrows, 
{\em Phys. Rev. Lett.} {\bf 106}, 161103 (2011).

\bibitem{reis4} ``Gravitational memory in binary black hole mergers'',
C. Reisswig and D. Pollney, 
{\em Astrophys. J. Lett.} {\bf 732}, L13 (2011).

\bibitem{ccespin} ``Gravitational-wave detectability of equal-mass black-hole
binaries with aligned spins'',
C. Reisswig, S. Husa, L. Rezzolla,  E.~N. Dorband, D. Pollney and J. Seiler,
{\em Phys. Rev. D} {\bf 80}, 124026 (2009).

\bibitem{extractool} ``A characteristic extraction tool for gravitational waveforms'',
M.~C. Babiuc, B. Szil\'{a}gyi, J. Winicour and Y. Zlochower.,
{\em Phys.Rev. D} {\bf 84}, 044057 (2011). 

\bibitem {einstk} http://www.einsteintoolkit.org.

\bibitem{isaac} ``Null cone computation of gravitational radiation'',
R.A. Isaacson, J.S. Welling and J. Winicour,
{\em J. Math. Phys.} {\bf 24} 1824 (1983).

\bibitem{highp} ``High-powered gravitational news'',
N.~T. Bishop, R.~G\'{o}mez, L. Lehner, M. Maharaj and
J. Winicour, {\em Phys. Rev. D} {\bf 56} 6298 (1997).

\bibitem{tam} ``Gravitational fields in finite and conformal Bondi frames'', 
L.~A. Tamburino and J. Winicour, {\em Phys. Rev.} {\bf 150} 1039, 1966.

\bibitem{bondi} ``Gravitational waves in general relativity VII.
Waves from axi-symmetric isolated systems'',
H. Bondi, M. van~der Burg and A.~W.~K. Metzner, 
{\em Proc. R. Soc. London A}  {\bf  269},  21  (1962).

\bibitem{sachs} ``Gravitational waves in general relativity VIII. Waves
in asymptotically flat space-time''', 
R.~K. Sachs,
{\em Proc. R. Soc.  London A} {\bf 270}, 103 (1962).

\bibitem{regwh} ``Stability of a Schwarzschild Singularity'',
T. Regge and J.~A. Wheeler, 
{\em Phys. Rev.} {\bf 108}, 1063 (1957).

\bibitem{zeril} ``Gravitational field of a particle falling in
a Schwarzschild geometry analyzed in tensor harmonics'',
F. Zerilli, 
{\em Phys. Rev. D.} {\bf 2}, 2141 (1970).

\bibitem{nagrez} ``Gauge-invariant non-spherical metric perturbations of
Schwarzschild black-hole spacetimes'',
A. Nagar and L. Rezzolla, 
{\em Class. Quantum Grav.} {\bf 22}, R167 (2005).

\bibitem{np} ``An approach to gravitational radiation by a method of spin coefficients'',
E.~T. Newman and R. Penrose,
{\em J Math. Phys.} {\bf 3}, 566 (1962).

\bibitem{bakcamluo} ``Modeling gravitational radiation from coalescing binary black holes'',
J. Baker, M. Campanelli, C.O. Lousto and R. Takahashi,
{\em Phys.Rev. D}, {\bf 65}, 124012--124034 (2002).

\bibitem{pretlet} ``Evolution of binary black-hole spacetimes'',
F. Pretorius, {\em Phys. Rev. Lett.} {\bf 95}, 121101 (2005).

\bibitem{camluomarzlo} ``Accurate evolutions of orbiting black-hole binaries
without excision'',
M. Campanelli, C. O. Lousto, P. Marronetti and Y. Zlochower,
{\em Phys.Rev.Lett.} {\bf 96} 111101 (2006).

\bibitem{bakcenchokopvme} ``Binary black hole merger dynamics and waveforms'',
J.~G. Baker, J. Centrella, D-I. Choi, M. Koppitz and J. van Meter,
{\em Phys.Rev. D} {\bf 73}, 104002 (2006).

\bibitem{lehnmor} ``Dealing with delicate issues in waveform calculations'',
L. Lehner and O.~M. Moreschi,
{\em Phys.Rev. D} {\bf 12}, 124040 (2007).

\bibitem{penrose} ``Asymptotic properties of fields and space-times'',
R. Penrose, {\em Phys. Rev. Lett.} {\bf 10}, 66 (1963).

\bibitem{winrev} ``Characteristic evolution and matching'',
J.  Winicour, {\em Living Rev. Rel.} {\bf 15}, 2 (2012).

\bibitem{helconf} ``Cauchy problems for the conformal vacuum field
equations in general relativity'', H. Friedrich,
{\em Commun. Math. Phys.} {\bf 91}, 445 (1983).

\bibitem{fraurev} ``Conformal infinity'',
J. Frauendiener,
{\em Living Rev. Rel.} {\bf 7}, 1 (2004).

\bibitem{saschrev1} ``Problems and successes in the numerical approach
to the conformal field equations'',
S. Husa,
{\em Lecture Notes Phys.} {\bf 604}, 239 (2002).

\bibitem{saschrev2} ``Numerical relativity with the conformal field equations'',
S. Husa,
{\em Lecture Notes Phys.}  {\bf 617}, 159 (2003).

\bibitem{zen} ``Hyperboloidal evolution with the Einstein equations'',
A. Zengino{\u g}lu,
{\em Class. Quantum Grav.} {\bf 25}, 195025 (2008).

\bibitem{moncrinn} ``Regularity of the Einstein equations at future null
infinity'',
V. Moncrief and O. Rinne,
{\em Class. Quantum Grav.} {\bf 26}, 125010 (2009).

\bibitem{rinn} ``An axisymmetric evolution code for the Einstein equations
on hyperboloidal slices'',
O. Rinne,
{\em Class. Quantum Grav.} {\bf 27}, 035014 (2010).

\bibitem{bardsarbuch} ``Tetrad formalism for numerical relativity on
conformally compactified constant mean curvature hypersurfaces'',
J.~M. Bardeen, O. Sarbach and L.~T. Buchman,
{\em Phys. Rev. D} {\bf 83}, 104045 (2011).

\bibitem{hordnull} ``General relativistic null-cone evolutions with a high-order
scheme''
C. Reisswig, N.~T. Bishop, and D, Pollney, arXiv:1208.3891 (2012).

\bibitem{ksachs} ``Observations in Cosmology"",
J. Kristian and R.~K. Sachs.,
{\em Astrophys. J.} {\bf 143}, 379 (1966).

\bibitem{ellis} `` Ideal observational cosmology'',
G.~F.~R. Ellis, S.~D. Nel, R. Maartens, W.~R. Stoeger, and A.~P. Whitman,
{\em Phys. Reports} {\bf 124}, 315 (1985).

\bibitem{bishc1}  ``Observational Cosmology and Numerical Relativity''
N.~T. Bishop and P. Haines,
{\it Quaest. Math.} {\bf 19}, 259 (1996).

\bibitem{bishc2} ``Observational cosmology using characteristic numerical
relativity'',
P. ~J. van der Walt and N. ~T. Bishop, 
{\it Phys. Rev. D} {\bf 82}, 084001 (2010).

\bibitem{bishc3} ``Observational cosmology using characteristic numerical relativity:
Characteristic formalism on null geodesics'',
P. ~J. van der Walt and N. ~T. Bishop, 
{\it Phys. Rev. D} {\bf 85}, 044016 (2012).

\bibitem{sachsdn} ``On the characteristic initial value problem in gravitational
theory'',
 R.~K. Sachs,
{\em J. Math. Phys.} {\bf 3}, 908 (1962).

\bibitem{close2} ``Retarded radiation from colliding black holes in the close
limit''
S. Husa, Y. Zlochower, R.  G{\'{o}}mez and J. Winicour,
{\em Phys.Rev. D} {\bf 65} 084034  (2002).
       
\bibitem{josh} ``Worldtube conservation laws for the null-timelike evolution
problem'',
J. Winicour, 
{\em Gen. Rel. Grav.}  {\bf 43}, 3269 (2011).

\bibitem{newtgc} ``Newtonian gravity on the null cone''
J. Winicour.
{\em J. Math. Phys.} {\bf 24}, 1193 (1983).

\bibitem{wobb} ``Moving black holes in 3D'',
R. G{\'{o}}mez, L. Lehner, R. Marsa and J. Winicour,
{\em Phys. Rev. D} {\bf 57}, 4778 (1998).
  
\bibitem{rend} ``Reduction of the characteristic initial value problem
to the Cauchy problem and Its applications to the Einstein equations'',
A.~D. Rendall,
{\em Proc. R. Soc. London A}, {\bf 427}, 221 (1990).

\bibitem{ccm} ``Cauchy-Characteristic Matching'',
 N.T. Bishop, R. G\'{o}mez, L. Lehner, B. Szil\'{a}gyi,
J. Winicour and R. A. Isaacson, in B. Iyer and B. Bhawal (Eds.), 
{\em Black Holes, Gravitational Radiation and the Universe}, 
Kluwer Academic Publishers, Dordrecht (1998).

\bibitem{holvorc2} ``Cauchy-characteristic evolution and waveforms'',
N.~T. Bishop, R. G{\'{o}}mez, P.~R.  Holvorcem, R.~A. Matzner, P. Papadopoulos
and J. Winicour, 
{\em J. Comput. Phys.} {\bf 136}, 140 (1997).

\bibitem{harl} ``Well-Posed Initial-Boundary Evolution in General Relativity''
B.  Szil\'{a}gyi and J. Winicour
{\em Phys.Rev.D} {\bf 68}, 041501 (2003).
    
\bibitem{wpgs} ``Problems which are well-posed in a generalized sense with
applications to the Einstein equations'',
H.O. Kreiss and J. Winicour,
{\em Class. Quantum Grav.} {\bf 23}, S405--S420 (2006).

\bibitem{isol} ``Boundary conditions for coupled quasilinear wave
equations with application to isolated systems'',
H-O. Kreiss, O. Reula, O. Sarbach and J. Winicour,
{\em Commun. Math. Phys.}  {\bf 289}, 1099 (2009).

\end{thebibliography}
\end{document}